# Timing jitter optimization of mode-locked Yb-fiber lasers toward the attosecond regime

**Youjian Song, Chur Kim, Kwangyun Jung, Hyoji Kim, and Jungwon Kim***

*KAIST Institute for Optical Science and Technology and School of Mechanical, Aerospace and Systems Engineering, Korea Advanced Institute of Science and Technology (KAIST), Daejeon 305-701, Korea*
*\*jungwon.kim@kaist.ac.kr*

**Abstract:** We demonstrate ultra-low timing jitter optical pulse trains from free-running, 80 MHz repetition rate, mode-locked Yb-fiber lasers. Timing jitter of various mode-locking conditions at close-to-zero intra-cavity dispersion (-0.004 to +0.002 $ps^2$ range at 1040 nm center wavelength) is characterized using a sub-20-attosecond-resolution balanced optical cross-correlation method. The measured lowest rms timing jitter is 175 attoseconds when integrated from 10 kHz to 40 MHz (Nyquist frequency) offset frequency range, which corresponds to the record-low timing jitter from free-running mode-locked fiber lasers so far. We also experimentally found the mode-locking conditions of fiber lasers where both ultra-low timing jitter and relative intensity noise can be achieved.

**OCIS codes:** (270.2500) Fluctuations, relaxations, and noise; (320.7090) Ultrafast lasers; (120.0120) Instrumentation, measurement, and metrology; (060.3510) Lasers, fiber.

## References and links

## 1. Introduction

Ultra-low timing jitter optical pulse trains from femtosecond mode-locked lasers enable extremely high timing-precision scientific and industrial applications such as long-range synchronization of accelerators, free-electron lasers (FELs) [1] and phased-array antennas [2], precise frequency comb generation [3], high-resolution optical sampling and analog-to-digital converters (ADC) [4], low-phase-noise microwave/RF signal generation [5-7], and coherent synthesis of optical pulses from multiple lasers [8,9]. Highly concentrated photon numbers in an ultrashort (e.g., <100 fs) laser pulse make the pulse temporal position robust against perturbations by photon noise, and it has been theoretically predicted that mode-locked lasers can achieve timing jitter well below a femtosecond [10-12].

Finding proper mode-locked lasers and mode-locking conditions that achieve minimal timing jitter is important for further advances in these high-precision applications. In doing so, it first requires the accurate timing jitter characterization of mode-locked lasers. Common timing jitter measurement methods based on direct photodetection of the pulse train and phase noise measurement of a selected RF harmonic [13] have limited dynamic range, which results in ~10 fs measurement resolution. To overcome this resolution limitation, a recently demonstrated balanced optical cross-correlation (BOC) method [14] can be used. The BOC method is an all-optical timing jitter characterization method that enables extremely high timing resolution (e.g., sub-20 as over the Nyquist frequency in this work) with minimal influence of intensity noise. The BOC method has recently been employed for timing jitter measurement of mode-locked solid-state and fiber lasers. Recent timing jitter measurements of solid-state Cr:LiSAF and Ti:sapphire lasers have shown 156 as and 20 as rms timing jitter, respectively, when integrated from 10 kHz to 10 MHz offset frequency range [15,16].

Femtosecond mode-locked fiber lasers [17] are attractive as ultralow-jitter signal sources because they are more compact, more robust, easier to build and operate, and lower-cost laser systems compared to solid-state crystal lasers. However, the fiber lasers are more challenging to optimize the noise performance than the solid-state lasers due to larger amplified spontaneous emission (ASE) noise, complicated pulse evolution dynamics in long fiber, limited pulse energy due to nonlinearities in fiber, and relatively lower cavity Q-factor. As a result, the best timing jitter performance demonstrated from free-running, passively mode-locked Er and Yb fiber lasers has been limited to the order of ~1 fs level so far [18-20].

Both the analytical theory [10,21] and numerical simulation [22] on the noise of mode-locked fiber lasers suggest that short pulsewidth and close-to-zero intra-cavity dispersion can reduce the timing jitter. The shorter pulsewidth can reduce the timing jitter that is directly induced by the ASE noise; the smaller intra-cavity dispersion can further reduce the timing jitter that is indirectly coupled by the center frequency fluctuations. In a recent study of mode-locked-regime-related timing jitter in Yb-fiber lasers, stretched-pulse regime indeed showed lower timing jitter than other mode-locked regimes such as soliton and self-similar owing to shorter pulsewidth and smaller intra-cavity dispersion magnitude [20]. However, due to relatively large chirp parameter at slightly positive net cavity dispersion condition (+0.003 $ps^2$ in [20]) typically used in stretched-pulse lasers, the measured rms timing jitter was limited to ~1 fs level when integrated from 10 kHz to 40 MHz (Nyquist frequency) offset frequency.

In this paper, we concentrate on the close-to-zero intra-cavity dispersion of free-running, passively mode-locked fiber lasers to search for the optimal timing jitter performance in fiber lasers. The experiments are based on stretched-pulse Yb-fiber lasers with grating pairs for intra-cavity dispersion compensation. The timing jitter of various mode-locking conditions at close-to-zero cavity dispersion (-0.004 to +0.002 $ps^2$ range at 1040 nm center wavelength in this work) is characterized using a sub-20-as resolution BOC method. The measured minimum rms timing jitter is 175 as when integrated from 10 kHz to 40 MHz (Nyquist frequency) offset frequency range. To our knowledge, this result corresponds to the lowest high-frequency timing jitter from mode-locked fiber lasers so far. In addition, the relative intensity noise (RIN) of different mode-locking conditions is characterized. We experimentally found the mode-locking conditions of fiber lasers where both ultra-low timing jitter (~200 as) and RIN ($<10^{-14}$ $Hz^{-1}$ at 10 kHz offset frequency) can be achieved.

## 2. Experimental setup

Two home-built, free-running, 80-MHz repetition-rate Yb-fiber lasers at 1040 nm center wavelength are used for the timing jitter characterization. The laser and experimental setup configuration is shown in Fig. 1. One laser is a unidirectional ring cavity, and the other laser is based on a σ-cavity design with a cavity mirror mounted on a piezoelectric transducer (PZT) for low-bandwidth repetition-rate locking with the other laser. The mode-locking operation in both lasers is achieved by nonlinear polarization evolution (NPE) in a stretched-pulse regime. A 600 line/mm grating pair is used in each laser for intracavity dispersion compensation. By tuning the separation of grating pairs, we can finely adjust the intra-cavity dispersion to search for the mode-locking condition that supports the lowest timing jitter. The total cavity normal dispersion is composed of single-mode-fiber, Yb-doped gain fiber and free-space bulk optics. The estimated cavity normal dispersion is +0.048 $ps^2$ at 1040 nm, which can be compensated by ~33 mm separation of the intracavity grating pair at 30-degree incident angle. The grating pair separation is tuned from 32 mm to 36 mm, where 1 mm grating pair separation change corresponds to a 2nd-order dispersion change of ~0.0015 $ps^2$ at 1040 nm. The intra-cavity net dispersion was measured using an in-situ dispersion measurement method [23,24], and the dispersion range is from +0.002(±0.001) $ps^2$ to -0.004(±0.001) $ps^2$ when the grating pair separation is tuned from 32 mm to 36 mm. Note that the cavity 2nd-order dispersion value changes significantly over the broad optical spectrum (>50 nm FWHM), which is mainly due to the high 3rd-order dispersion introduced by the grating pair. In this paper, the quoted intra-cavity dispersion is the 2nd-order dispersion value at a nominal center wavelength of 1040 nm for all the measurements.

The timing jitter of Yb-fiber lasers is measured by the BOC method. A SF11 prism pair is used for extra-cavity dispersion compensation of each laser before the cross-correlation in order to enhance the resolution for the BOC-based timing jitter characterization. The BOC method measures the changes in temporal overlap between two optical pulses with minimal influences of intensity noise, which enables sub-20-attosecond-resolution detection of timing jitter between optical pulse trains. More detailed information on the BOC measurement technique can be found in [14] and [20]. By monitoring the jitter spectral density from the BOC using RF analyzer and FFT analyzer, we search for the optimal dispersion and mode-locking conditions that achieve the lowest timing jitter.

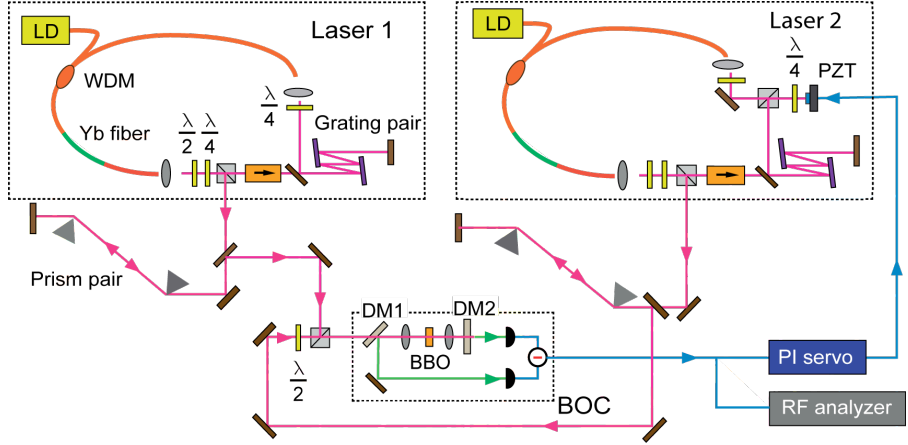

Fig. 1 Experimental setup of the BOC-based timing jitter characterization of Yb-fiber lasers. DM: dichroic mirror; LD: 976nm laser diode; PZT: piezoelectric transducer.

## 3. Timing jitter characterization at close-to-zero intra-cavity dispersion

The timing jitter spectra at various mode-locking conditions in the stretched-pulse regime are characterized using the BOC method. The typical timing jitter spectra and optical spectra from -0.004 $ps^2$ to +0.001 $ps^2$ range are plotted in Fig. 2. For comparison, the jitter and optical spectrum at +0.003 $ps^2$ measured in [20] is also shown. The timing jitter spectra follow $1/f^2$ slope from 10 kHz to 1 MHz offset frequency, which indicates the random walk nature directly originated from the ASE noise-induced timing jitter. Interestingly, when the intra-cavity dispersion approaches zero, the jitter spectral density can be 18 dB lower than the same stretched-pulse laser working at +0.003 $ps^2$ intra-cavity dispersion.

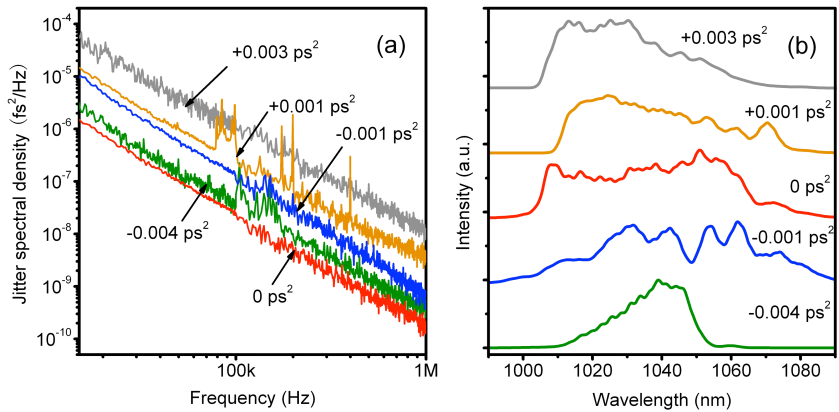

Fig. 2. (a) Typical timing jitter spectra and (b) the corresponding optical spectra measured at different mode-locking conditions from a stretched-pulse Yb-fiber laser.

The integrated timing jitter from 10 kHz to 10 MHz versus intra-cavity dispersion for different laser mode-locking conditions is plotted in Fig. 3. The lowest integrated timing jitter can be achieved at zero intra-cavity dispersion. There is asymmetric increase of integrated timing jitter versus intra-cavity dispersion, which is due to different chirp parameter for positive and negative dispersion. In stretched-pulse fiber lasers, the Kerr effect and dispersion imbalance results in intracavity pulse chirping. The chirp parameter satisfies the following equation according to S. Namiki and H. A. Haus's analytical theory [21],

$$\beta = \tan\left\{\frac{1}{2}\left[\arg(\alpha - j) - \arg\left(\frac{g}{\Omega_g^2} + jD\right)\right]\right\} \quad (1)$$

where $g$ is laser amplitude gain, $\Omega_g$ is HWHM of gain bandwidth of laser medium, $D$ is intra-cavity dispersion, and $\alpha$ is proportionality factor that depends on the orientation of wave plates and polarizers. For our Yb-fiber lasers, $\Omega_g$ is $3.9\times10^{13}$ rad/s; $g$ is determined by compensating the cavity loss, which is calculated as ~1; $\alpha$ is between 0.1 to 0.3 for typical NPE mode-locking fiber lasers [21] and is set to 0.2 in this paper. The calculated chirp parameter versus cavity dispersion is also plotted in Fig. 3. The chirp parameter is nearly zero for negative dispersion, and dramatically increases in magnitude at positive dispersion as shown in Fig. 3. The larger absolute value of chirp parameter results in a longer average intra-cavity pulse duration, which leads to larger timing jitter. This explains the rapid increase of timing jitter at positive intra-cavity dispersion. Note that the chirp parameter is non-zero at zero intra-cavity dispersion. This explains the reason why the integrated timing jitter at zero dispersion is not much reduced compared to the jitter at the slightly negative dispersion (e.g., at -0.004 ps$^2$ in Fig. 3) even though the indirectly-coupled timing jitter is minimized at zero cavity dispersion.

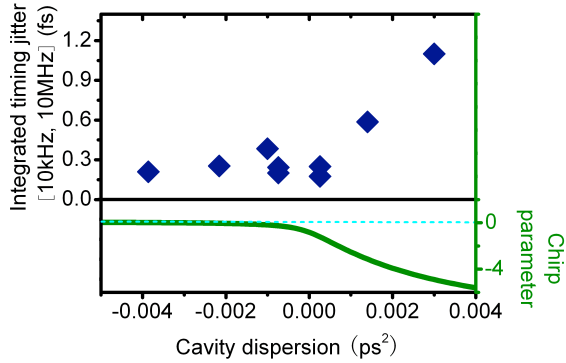

Fig. 3. The rms timing jitter (integrated from 10 kHz to 10 MHz offset frequency) versus intra-cavity dispersion of different mode-locking conditions. Bottom is the calculated chirp parameter versus intra-cavity dispersion.

The timing jitter spectral density and the equivalent single-sideband (SSB) phase noise at 10-GHz carrier frequency of the lowest integrated timing jitter condition is shown in Fig. 4. The net cavity dispersion is 0.000(±0.001) ps$^2$. The inset shows the optical spectra of two lasers with FWHM of ~55 nm. The shot noise level (~$10^{-12}$ fs$^2$/Hz) is lower than the measured results over the entire Nyquist frequency, which indicates that the measurement is not limited by the BOC resolution. The rms timing jitter integrated from 10 kHz to 40 MHz (Nyquist frequency) offset frequency is 175 as (shown in the bottom of Fig. 4). To our knowledge, this is the lowest high-frequency timing jitter performance measured from mode-locked fiber lasers. The flat jitter spectrum above 7 MHz offset frequency can be explained by the RIN-coupled timing jitter originated from the Kramers-Krönig relation [11]: the resulting timing jitter spectral density can be expressed by $S_{\Delta t^2}(f)=RIN(f)/(2\pi\Delta f_g)^2$, where $\Delta f_g$ is the gain bandwidth. By using the measured laser RIN of ~$10^{-14}$/Hz and the gain bandwidth of 45 nm, we can predict the RIN-induced timing jitter spectral density to be ~$2\times10^{-12}$ fs$^2$/Hz level, which agrees fairly well with the measured result. For comparison, the RIN-induced timing jitter projected from the measured RIN is also plotted in Fig. 4.

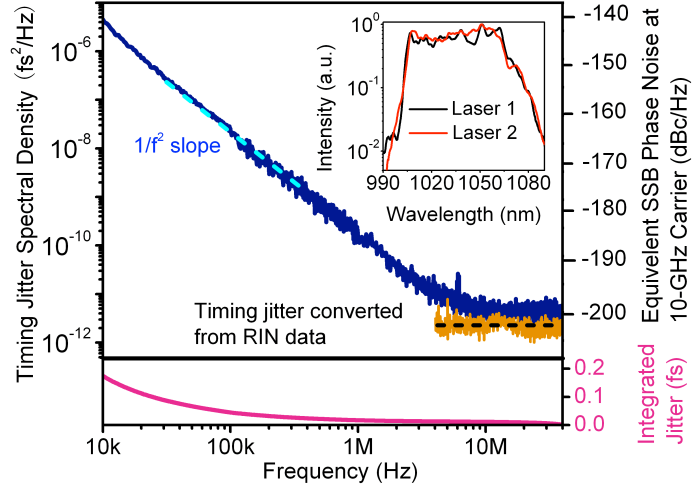

Fig. 4. Top: The best timing jitter spectral density measurement result and the equivalent single-sideband (SSB) phase noise at 10-GHz carrier frequency of the Yb-fiber laser operating at zero intra-cavity dispersion. The RIN-induced timing jitter projected from the measured RIN is also plotted. Bottom: The integrated timing jitter is 175 as [10 kHz – 40 MHz]. Inset: optical spectra of the two Yb-fiber lasers used.

## 4. Comparison of timing jitter and RIN performances at close-to-zero cavity dispersion

As shown in Fig. 3, mode-locking conditions at the negative dispersion side of the close-to-zero cavity dispersion can support sub-500 as timing jitter. Several mode-locking conditions can achieve ~200 as timing jitter performance besides the 175 as integrated timing jitter at the zero cavity dispersion. At a fixed cavity dispersion, different mode-locking conditions can have more than twice difference in the integrated timing jitter value. In this section, we compare two interesting cases for mode-locking condition dependent timing jitter.

The timing jitter spectra measured at -0.004 $ps^2$ and 0.000 $ps^2$ cavity dispersion conditions are plotted in Fig. 5. The corresponding integrated timing jitter is marked as red up-triangle (at -0.004 $ps^2$) and blue down-triangle (at 0.000 $ps^2$) in the upper inset of Fig. 5. The FWHM of the output optical spectrum at the zero cavity dispersion is more than twice wider than that of the negative cavity dispersion condition, as shown in the lower inset of Fig. 5. However, the BOC measurement shows that the timing jitter spectra and the resulting integrated jitter (175 as for 0.000 $ps^2$ and 210 as for -0.004 $ps^2$) are very similar even though the dispersion condition and output optical spectra are quite different. This result shows that it is not a necessary condition to operate the fiber laser at the exact zero dispersion in order to achieve ~200 as timing jitter. In the negative cavity dispersion, the reduced chirp parameter enables a low timing jitter performance, even though its indirectly-coupled timing jitter is larger than that of the zero cavity dispersion condition. When we increase the grating pair separation further to get a larger negative cavity dispersion (<-0.004 $ps^2$), the timing jitter increases again because the average pulse duration now becomes significantly longer.

When the cavity dispersion of mode-locked fiber lasers approaches zero, the mode-locking condition can be quite different by changing the NPE strength and finely tuning the grating pair separation. As a result, not all the mode-locking conditions guarantee a sub-200 as timing jitter. For some mode-locking conditions, even when the RF spectrum, optical spectrum, and extra-cavity dechirped pulsewidth are similar, the jitter spectrum can differ much. This is the case as shown by the green up-triangle (at -0.001 $ps^2$) and the blue down-triangle (at 0.000 $ps^2$) of the upper inset in Fig. 6. The jitter spectra and optical spectra of these two mode-locking conditions are shown in main part and the lower inset of Fig. 6, respectively. Even the optical spectra look very similar with almost identical dispersion conditions, the integrated timing jitter at -0.001 ps$^2$ in a non-optimal laser condition (390 as) is more than twice larger than that of the best timing jitter at 0.000 ps$^2$ (175 as). This also shows the usefulness of the BOC method that it can serve as an ultra-sensitive timing jitter status monitor, which is necessary for maintaining the fiber lasers in the minimum jitter condition for noise-sensitive applications.

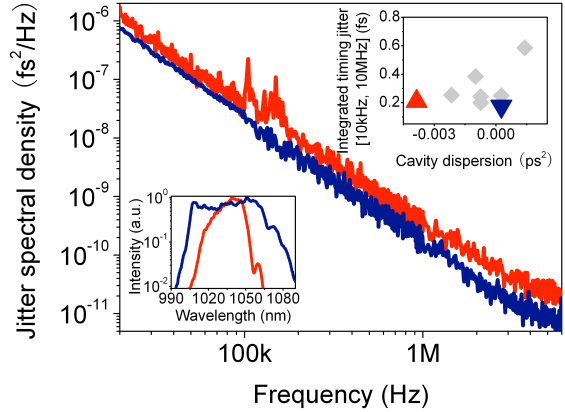

Fig. 5. The comparison of best achievable timing jitter at 0 ps$^2$ (blue down-triangle and blue curves) and -0.004 ps$^2$ (red up-triangle and red curves) intra-cavity dispersion. Even the dispersion and optical spectra are different, the jitter spectra and integrated jitter are similar.

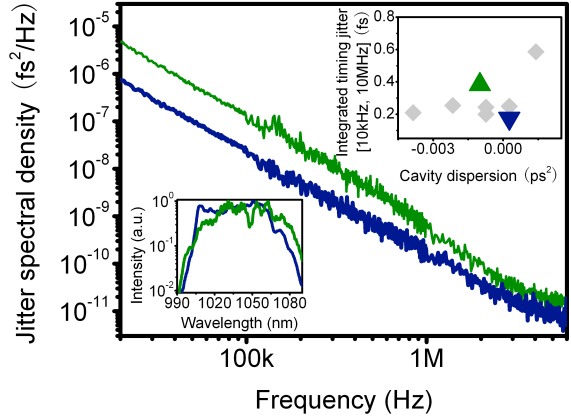

Fig. 6. The comparison of timing jitter with different mode-locking conditions. Even the dispersion and optical spectra are similar, the jitter spectra and integrated jitter can be significantly different.

Recently, the RIN of mode-locked fiber lasers has been discussed intensively [24-27]. In this work, in addition to the timing jitter characterization, we also measured the RIN of the above-mentioned mode-locking conditions. Fig. 7 shows the measured RIN spectra of various mode-locking conditions of the Yb-fiber laser. The RIN data measured at -0.021 ps$^2$ cavity dispersion corresponds to the soliton regime. The other data are measured in the stretched-pulse regime. The RIN of mode-locked fiber lasers obtained at zero cavity dispersion and negative cavity dispersion is lower than that of positive cavity dispersion in the high offset frequency (>30 kHz), which is similar to the timing jitter measurement results shown in Fig. 3. The lowest RIN can be obtained at close-to-zero intra-cavity dispersion conditions (-0.004 ps$^2$ to 0 ps$^2$ range), which is consistent with the recent study in [24]. The lower RIN at slightly negative dispersion might be due to the soliton-like pulse formation effect in negative dispersion as explained in [28].

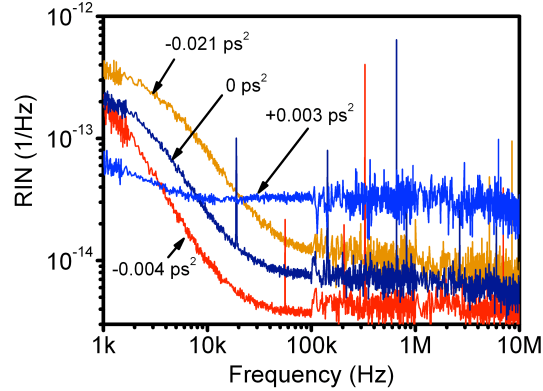

Fig. 7. RIN of the mode-locked Yb-fiber laser with different intra-cavity dispersion.

## 5. Conclusion and discussion

In this paper, we characterized the high-frequency timing jitter and RIN of free-running, stretched-pulse Yb-fiber lasers operating at close-to-zero intra-cavity dispersion. The measured lowest rms timing jitter is 175 as when integrated from 10 kHz to 40 MHz offset frequency. To our knowledge, this result corresponds to the lowest high-frequency timing jitter from mode-locked fiber lasers so far. This result demonstrates that standard free-running, NPE-based fiber lasers can achieve timing jitter (and equivalent phase noise) performance comparable to solid-state crystal lasers [15,16] and the best commercial microwave sources (such as sapphire-loaded cavity oscillators) with much reduced cost and engineering complexity. Another interesting finding is that both the lowest timing jitter and RIN can be obtained in a narrow range of close-to-zero dispersion (in this work, from -0.004 $ps^2$ to 0 $ps^2$), which is fairly consistent with the recent study on the optimization of $f_{ceo}$ noise at zero dispersion [24]. Since choosing the right mode-locking condition at a given intra-cavity dispersion is also important for the optimization of timing jitter, the BOC method can be used as an ultra-sensitive, real-time jitter monitor to find and maintain the best performance. Note that the Yb-fiber laser used in this work is not fully optimized for the lowest possible timing jitter operation because of the low cavity Q (four bounces on grating pair in one round-trip contribute 85 % power loss). Higher Q fiber lasers (e.g., all-fiber implementation) operating at close-to-zero cavity dispersion are expected to have timing jitter well below 100 as in the near future.

## Acknowledgement

This research was supported by the National Research Foundation of Korea (NRF) funded by the Ministry of Education, Science and Technology (MEST, 2010-0003974).